# Feeding the human brain model

Paul Tiesinga[1], Rembrandt Bakker[1,2], Sean Hill[3], and Jan G. Bjaalie[4]

[1]Donders Institute, Radboud University Nijmegen, Heyendaalseweg 135, 6525 AJ Nijmegen, The Netherlands
[2]Institute of Neuroscience and Medicine (INM-6) and Institute for Advanced Simulation (IAS-6), Jülich Research Centre and JARA, 52425 Jülich, Germany
[3]Blue Brain Project, Campus Biotech, Bâtiment B1, Ch. Des Mines 9, Geneva CH-1202, Switzerland
[4]Institute of Basic Medical Sciences, University of Oslo, PO Box 1110 Blindern N-0317 Oslo, Norway

**Abstract**
The goal of the Human Brain Project is to develop during the next decade an infrastructure necessary for running a simulation of the entire human brain constrained by current experimental data. One of the key issues is therefore to integrate and make accessible the experimental data necessary to constrain and fully specify this model. The required data covers many different spatial scales, ranging from the molecular scale to the whole brain and these data are obtained using a variety of techniques whose measurements may not be directly comparable. Furthermore, these data are incomplete, and will remain so at least for the coming decade. Here we review new neuroinformatics techniques that need to be developed and applied to address these issues.

**Introduction and background**

A key goal of the 1 billion euro, 10 year Human Brain Project (HBP) is to build a scaffold model of the human brain. This will enable the global community iteratively build and refine whole brain models, starting from the mouse and working towards the human brain, which is about a thousand times larger. Different teams of researchers will each deal with different sets of challenges. One set of challenges is to develop the hardware and software to make it possible to simulate such a large-scale model, store and analyze its output, and control the simulation. Another set of challenges is to fully specify the computational model that needs to be simulated, and identify key missing data, which is the topic of this review. When these challenges are met, the HBP infrastructure will provide the community with new tools to accelerate the understanding of the brain in health and disease. The HBP approach can be seen as neuroscience data integration through model building. The idea is that only by building an integrated model will neuroscientists be able to find out whether there is missing data and, if so, determine an experimental strategy to measure it directly or use a predictive neuroinformatics approach (see below) to infer it.

Simulation of the brain models will occur at different scales and levels of abstraction. The cellular-level biophysical simulation is in simplified terms the numerical integration of a set of coupled (partial) differential equations. These equations, their parameters, and their initial conditions, need to be fully specified to run the simulation. For instance, there are differential equations that describe the time evolution of the membrane potential and ionic concentrations as a function of spatial location within a neuron [1,2]. Other equations describe the molecular cascades inside the cell [3], such as those involving transcription and translation into proteins [4,5] that are, for instance, necessary to lay down memories [6].

The general form of the equations is well established, and even though there may be discussion on what level of simplification is acceptable, the main challenge is to populate the model parameters with reasonable estimates.

To obtain the necessary data a three-pronged strategy is needed. **First**, integrate existing data from different labs. **Second**, predict missing data that have not (yet) been measured. **Third**, increase the amount of available experimental data through new molecular neurobiology techniques and industrial neuroscience approaches (i.e. high throughput). A key example of this approach is the work by the Allen Institute for Brain Science (AIBS).

**The data integration challenge**
The information constraining the HBP model comes from diverse sources and is obtained using different experimental techniques. Hence, for the same basic assertion, say the likelihood of a connection between two neurons in area A and B, there are multiple sources of data, each potentially giving a different answer. These data need to be integrated. A key problem is the representation of information in such a way that they are comparable and so that their reliability and precision can be quantified and taken into account.

A group of researchers working in the same lab knows the experimental settings used when measurements are made. These measurements are stored in data files, most likely without documentation of what the units are, or what preprocessing (filtering) has been done. Information regarding sampling rates and other parameters might be stored separately (for instance, written in a lab book) or assumed to be the default setting specified in the protocol. The data will be subsequently analyzed using tools bundled with the experimental equipment or lab-specific user-written code. It will be stored in another file, without explicitly mentioning these post-processing steps and their parameters. This set up makes sense within the lab, but it means that the data are practically useless outside the lab, because the metadata – the additional information about the experiments – is missing. Further, standardization of the metadata is required to assure correct interpretation. Hence, an ontology of the concepts describing the data and measurement parameters needs to be developed. There are ontologies for neuroscience concepts, collected in the Neuroscience Information Framework (NIF) [7], but they do not yet cover the necessary experimental and analysis protocols in a comprehensive way. When, for instance, axonal projection patterns are determined using tracing experiments, the protocol can be structured in different modules, each of which can be characterized by well-defined parameters out of an ontology or controlled vocabulary (Figure 1A). When the data are subsequently stored in a database and referenced relative to a standard atlas, this information can be easily accessed by other researchers [8] (Figure 1B).

**Predictive neuroinformatics**
Predictive neuroinformatics aims to fill in missing data based on existing data and general principles. It naturally builds on methods developed in other fields where similar problems have been encountered. For instance, during clinical trials, for a given subject, sometimes only a part of the measurements are conducted. This leads to incomplete data, with missing entries for particular subjects, which need to be filled out using so called imputation techniques [9]. One can consider these data as a matrix, where each row represents a subject and each column a feature. Microarray experiments can also be expressed as a matrix, where the rows represent conditions and the columns the expression level for each gene [10], or for genome wide association studies (GWAS) where the genotypes of a single nucleotide polymorphism (SNP) are placed in a matrix, with the row representing the subject and the column the SNP location [11]. The empty elements in matrix can be filled by replacing each empty element by the mean value of the corresponding non-empty elements, but more advanced techniques are available [10,12,13]. A recent approach, directly applicable to neuroscience, is matrix completion [14,15]. This approach assumes that the data matrix is comprised of a sum of a low-rank matrix and a sparse matrix [16]. The solution for this problem resembles the well-known LASSO procedure for regression [17], in which an L1 (i.e. absolute value) penalty term on the regression weights is added to the L2 loss function (i.e. squared difference). This shrinks the weights to zero, with those smaller than a specific value made exactly zero. In the matrix version, the low rank part is found by applying a singular value decomposition (SVD) and shrinking the singular values to obtain, after reconstitution of the components with the new singular values and after a number of iterations, the best fitting low rank matrix [18,19].

The wiring diagram of the nervous system in C. Elegans has been determined [20,21], and the gene expression profile of each cell has also been measured and made publically available (http://www.wormbase.org). Analyses reveal that the expression profile can successfully predict the absence or presence of a synapse [22,23]. For the mouse, gene expression patterns [24] and meso-scale connectivity [25] have been made publicly available by AIBS. (Note: it would be more appropriate to refer to the AIBS meso-connectome as macro-scale connectivity because currently it is at the level of brain areas rather than cell populations.) When a similar analysis was applied at the level of brain areas rather than individual cells, it revealed that the connectivity pattern was to a large extent predictable by areal patterns of expression [26,27]. This means that when gene expression patterns are available, but connectivity data are not, the connectivity can still be predicted.

**Experimental sources of data to constrain the model**

**Brain areas.**
To build the brain model, the different brain areas need to be defined (a parcellation), their typical size determined, as well as the density of each cell type within that area and the distribution across substructures (layers, nuclei). A number of brain parcellations have been proposed, each based on different criteria, such as cytoarchitecture or the density with which receptors are expressed [28]. Recently, progress has been made with data-driven approaches to define brain areas. As mentioned above, AIBS has made available an atlas of transcription patterns in the mouse brain [24]. Each voxel was characterized by a vector, which contained the strength of expression of each of the analyzed genes. Nearby voxels, likely belonging to the same area, should also have a similar expression pattern. When a clustering procedure was applied to these vectors a parcellation emerges [29] (Figure 2D), which when 60 or more clusters are sought, resembles parcellations based on cytoarchitecture in microscopic sections (Nissl stains) or MRI contrast [30,31] (Figure 2E).

Each brain area is characterized by its shape, size, the cell types it contains and their density. Cell density is only known for a few well-studied areas such as the barrel cortex [32], but it can also be inferred as follows. The transcription pattern for a voxel is the weighted sum of the transcription pattern of the cell types that are present in that area. Hence given the individual cell-type patterns [33,34], the relative density of each cell type can be reconstructed [35]. The data so obtained matches the expected location of the cell types considered, i.e. cerebellar granular cells are recovered in the cerebellum. Layer-specific transcriptions patterns have also been obtained [36] making it possible to also infer cell-specific densities for each layer. These 'big data' analyses, in which different publicly available data sets are combined to infer unmeasured quantities are valuable for constraining whole brain models.

**Connections between neurons**
The cell-based data is manageable because it can be obtained locally, that is, by fully analyzing a particular area of the brain, independently of other areas, or by patching, filling and reconstructing the morphology of a single cell of that type one at a time. Connectivity, can not be dealt with in that way, because the identity of the pre- and postsynaptic cells needs to be known, the corresponding cell bodies of which are often in different areas, sometimes far from each other. Furthermore, the axonal/dendritic branch on which the synapse is made needs to be known to properly account for effects of postsynaptic integration [37]. Techniques to estimate connectivity are currently undergoing rapid development.

The classical approach for estimating connectivity at the microcircuit level is to record from two or more cells simultaneously in a slice, and repeat this across many slices to sufficiently sample the network, so that various statistics, such as the connection probability between different cell types and its distance dependence can be accurately determined [38-40]. When the cells are filled and subsequently reconstructed [41], information about the branch statistics of synapses can also be obtained [42]. In more recent approaches, in which cells are labeled by fluorescent proteins and the whole brain is imaged using an automated fluorescence micro-optical sectioning tomography system, a much larger number of cells, including axonal projections, can be reconstructed [43] (Figure 3B).

Molecular neurobiology is making great strides methodologically [44]. For example, Kim and coworkers have recently adapted a technique used in drosophila for use in mammalian brains [45]. A green fluorescent protein (GFP) is split into two parts and attached to two other proteins, one that is expressed presynaptically and another that is expressed postsynaptically. When at some location a pre- and postsynaptic side are a small distance apart, and each express the appropriate part of the protein, the GFP is reconstituted and becomes fluorescent [46] (Figure 3A). This method, referred to as GRASP was applied to determine the statistics of the CA3 to CA1 projection in hippocampus [47] (see below).

Improvements have also been made in determining macro- and meso-scale connectivity, which is the connectivity between brain areas and cell populations, respectively (Figure 2B and C). The typical approach was to bulk inject a dye, which would then either anterogradely label the axonal projections from the injection site, or retrogradely label the cell bodies of the neurons that project to the injection site [48]. AIBS has standardized the protocols for performing this type of experiments. AIBS researchers inject at various locations in the brain a virus that expresses GFP in the cells near the injection site and subject the labeled brain slices to a standardized analysis protocol [25,49] (Figure 2A). Although it serves as a proof of principle, their analysis does not give enough information to reconstruct a connection matrix at the cell-to-cell resolution. Specifically, the strength of connection is quantified as the number of subvoxels in a 100 μm voxel in the target area, for which the fluorescent signal exceeds a threshold (Figure 2B). This measure therefore does not directly translate in the number of synapses made in the target area, which would be the quantity of interest. Further, terminal fields of fiber and passing fibers cannot be easily distinguished from each other. Another (non-viral) approach is to co-inject two pairs of anterograde and retrograde tracers in two areas, say area A and B. In this way, four different connections can be discovered, including an indirect connection via an intermediate area [50], which yields similar connection matrices (Figure 2C). The next step is to use transgenic animals expressing Cre in a specific cell type [51,52] and inject a virus with a gene for a fluorescent protein that is floxed. Using this virus, the projection from specific cells types, which are sometimes restricted to a specific cortical layer or subcortical nucleus, can be anterogradely labeled. In a recent study the inputs to specific serotonergic neurons in the raphe nucleus were retrogradely traced using modified rabies viruses [53] (Figure 3C), which yields information about connectivity at cellular resolution.

A simple geometric principle, Peters' rule [54,55], can predict whether there is a potential synapse between two neurons. A synapse can only be present when an axon and a dendrite come close enough for the spine (for excitatory synapses) to reach the presynaptic bouton, which leads to a method to construct the connectivity based on the morphology of the postsynaptic dendrite and the presynaptic axon. The key question is what information can be reliably extracted from such a construction. One can estimate the number of synapses between the different populations [56], ignoring between which neurons the synapses are, or determine the resulting firing rates resulting from the numbers obtained in this way [57]. One can also estimate the distribution of synapses across branch order or path length from soma [42,58]. Overall, these measurements are consistent with experimental data, because they are aggregated and represent the properties of the density of synaptic elements. Peters' rule also makes some predictions for the synaptic density at the level of individual neurons. For instance, if the length of axonal branches in the neighborhood of the dendrite is constant, synapses from that source should have a uniform density. This means that the number of synapses is proportional to the length of the dendritic branch and the distances between two consecutive synapses on a branch are exponentially distributed. Data obtained using the GRASP method shows that for the CA3 to CA1 projection both of these statistics deviate from the predictions made by Peters' rule [47] (Figure 3A).

Although at the aggregate level Peters' rule predicts the number and location of synapses, it does not do so at the cell-to-cell level. This suggests that connectivity matrix constructed from the density representation of morphologies will match the average properties of a real connectivity matrix, such as degree distribution, but will deviate from higher order properties. This presents two issues for building brain connectivity. First, probabilistic models for non-random selection of potential synapses need to be developed. These should at least incorporate cell-type specific rules, because glutamate-uncaging experiments have shown [59] that the strength of a connection can be predicted from cell morphology, but with a proportionality constant that depends on the identity of the projection. This likely also holds for inhibitory circuits [60,61]. Second, they could incorporate additional labels that specify cell identity at sub cell-type granularity.

This review has focused primarily on the renaissance of anatomical studies that is providing detailed structural information of neural circuits at cellular resolution. An alternative strategy for characterizing neural circuits is to record neural activity using imaging methods or electrodes, as reviewed in the other papers in this issue, and to infer connectivity by various means [62-64]. So far these approaches have mostly focused on local circuits [65-67] or when long-range projections are targeted [68], the resolution is low, in essence similar to the meso-scale connectome discussed in preceding text [25]. Long-range projections at the cellular resolution are now coming into sight with new imaging technology that can cover two brain areas simultaneously [69] and with flexible electrodes on the brain surface that can record spikes of individual neurons [70,71]. The structural data will be used in the HBP to simulate the neural activity patterns, which link spiking activity at the single neuron level to large-scale activity at the level of brain areas. The emerging large-scale recording techniques will therefore be essential to validate these simulations.

Taken together, it is expected that through rapid development and subsequent adoption of new molecular neurobiology approaches more and more high quality connectivity data will become available, which places strong demands on methods to integrate these different types of data. HBP can make a contribution to these needs by providing a platform for data integration and the development of new predictive neuroinformatics methods.

**Acknowledgments**. The research leading to these results has received funding from the European Union Seventh Framework Programme (FP7/2007-2013) under grant agreement n° 604102 (HBP) and 600925 (NEUROSEEKER, PT). PT was also supported by the Netherlands Organisation for Scientific Research (NWO), through a grant entitled "Reverse physiology of the cortical microcircuit," grant number 635.100.023 and the Netherlands eScience Center through grant 027.011.304 (Biomarker Boosting). We thank Marijn Martens for comments on the manuscript.

## Figure captions

**Figure 1**. *Data needs to be properly organized and described in order to be accessible and interpretable for the scientific community.* Detailed knowledge about the anatomical organization of axonal connections is important for understanding normal functions of brain systems and disease-related dysfunctions. To explore more efficient ways of mapping, analyzing, and sharing detailed axonal connectivity data from the rodent brain, a workflow for data production and an atlas system tailored for online presentation of axonal tracing data is necessary. The Rodent Brain WorkBench (www.rbwb.org) is an example of such a tool. (A) Flowchart showing four processing steps, beginning with an animal submitted to an axonal tract-tracing experiment, followed by tissue and image processing steps, to an end result consisting of section images in a database. For each step, information about experimental parameters and procedures (metadata) are stored together with the section images. (B) Diagram showing module details and the input and output elements of each module in the workflow. Figure is adapted from [8].

**Figure 2**. *Determination of meso-scale connectivity requires robust atlasing technology.* (A) A virus was injected at a particular location and labeled the dendrites and axons of the cells it infected. Fluorescence highlights both the targets of the axonal projection as well as the fibers going there [25]. To identify which areas are the targets of projections, (left) the fluorescent images need to be mapped onto (middle) slices from a reference brain through a registration procedure, so that the identity of the labeled areas can be determined in a 3D reference model. (B) The strength of each projection, quantified as the degree of fluorescence, can then be represented in a connection matrix [25]. (C) Similar connection matrix restricted to cortical connections was obtained using non-viral tracers [50]. (D) Underlying this matrix is a parcellation of the brain areas. This can be data driven or done by manual segmentation. For instance, the gene expression pattern can be determined for each voxel, and then the voxels can be clustered in different groups based on their expression patterns [29]. When each cluster gets a different color, a realistic looking

parcellation emerges. Left: using 50 clusters and right: using 60 clusters. (E) Parcellation based on MRI data and expert annotation [31]. Panel A and B are reprinted by permission from Macmillan Publishers Ltd: Nature [25], copyright 2014; C is reprinted by permission from [50]; D is reprinted from [29] and E is reprinted by permission from [31].

**Figure 3**. *New techniques make it possible to sample synaptic connections at the cellular level.* (A) GFP Reconstitution Across Synaptic Partners (GRASP). (1) Cells express GFP fragments localized either to the presynaptic or to the postsynaptic extracellular spaces, respectively. At synapses between the two cells the split GFP fragments interact to reconstitute GFP. (2) Dendrite showing a number of reconstituted GRASP signals (green) in sites where dense CA3 axons (blue) intersect with a CA1 dendrite (red) (left) and its reconstruction showing the detected GRASP hot spots (right). (3) Synaptic density of CA1 dendrites with respect to control Poisson model taking into account axon density of the CA3 projection. Branches in blue and red show significantly less or more synaptic density than predicted, respectively (left). Heat-map of axonal density surrounding the branches (right). (B) Reconstruction of long-distance projection neurons in a mouse brain. High-resolution three-dimensional volume rendering of the local brain region, including the somatosensory cortex, with the four neurons that were reconstructed labeled by Roman numerals (i, ii, iii, and iv). Scale bars=500 μm. (C) Monosynaptic inputs that target serotonergic neurons in the dorsal raphe (DR) were retrogradely traced using a modified rabies virus. 3D visualization in one mouse showing 73,724 EGFP-labeled cells (green dots) from top (left) and angle (right) view. Red dots are the starter population in DR. Panel A1 is reprinted by permission from [72], A2 and A3 from [47]; B is reprinted by permission from [43]; C is reprinted by permission from [53].

## Key papers

**[25] Uses a systematic experimental approach together with an informatics pipeline to publish the first, more or less complete, meso-scale mouse connectome.

**[47] Applies the GRASP method to characterize the CA3 projection to CA1 and finds that synapse placement on some synaptic branches does not conform to the predictions of Peters' rule.

**[29] Re-analyzed the AIBS gene expression atlas (Ref [24]) to produce parcellations of the mouse brain based on similarity between gene expression patterns of each pair of voxels, which match manual parcellations based on anatomical considerations.

**[35] Combined the AIBS gene expression atlas (Ref [24]) with single neuron gene expression patterns to infer the fraction of each cell type in a given brain area.

*[41] Describe resources for analyzing and modeling the morphology of neurons and shows the impact data sharing can have in producing better models and improving analysis methodologies.

**[40] Painstaking experiments to obtain data that characterizes a novel circuit generated by two interneuron types residing in layer 1: One that has a broad horizontal impact and another that controls inhibition within a column.

**[61] Reports on details of the interconnectivity of different interneuron types in cortex.

**[43] A new imaging technique fMOST (Ref [73]) is applied to trace the axonal projections of a few neurons throughout the entire brain.

*[74] Report on an initiative to measure human brain connectivity non-invasively at high resolution.

**[42] The authors compare the branch statistics of synaptic locations obtained by light microscopy to those predicted from Peters' principle applied to large sets of replicated neurons based on experimentally reconstructed morphologies.

*[33] Transcriptomics of single neurons reveals differences in gene expression profiles of cell types.

**[50] Report on cortical connectivity obtained through co-injection of two sets of pairs of anterograde and retrograde tracers. Reveals mouse connectome similar to that of [25] but restricted to cortex.

**[49] Use injections of AAV virus in thalamus to determine pattern of thalamocortical projections. Identifies the thalamic nuclei that target specific cortical areas and provides a layer-resolved thalamic projection pattern for vibrissal motor cortex.

**[53] The origin of the synaptic inputs to serotonergic neurons in the Raphe nucleus is determined using a modified rabies virus.

*[44] Review of recent molecular methods, including those for determining neural connectivity.